\documentclass[nofootinbib,floats,aps,superscriptaddress,preprint]{revtex4}

\usepackage{graphicx}
\usepackage[hypertex]{hyperref}

\newcommand{\beqa}{\begin{eqnarray}} 
\newcommand{\eeqa}{\end{eqnarray}} 
\newcommand{\beq}{\begin{equation}} 
\newcommand{\eeq}{\end{equation}} 

\newcommand{\lsim}{\lesssim}
\newcommand{\gsim}{\gtrsim}

\def\TeV{\mbox{TeV}}
\def\GeV{\mbox{GeV}}
\def\MeV{\mbox{MeV}}
\def\eV{\mbox{eV}}
\def\keV{\mbox{keV}}

\begin{document}

\title{Natural soft leptogenesis}

\author{Yuval Grossman}
\affiliation{Department of Physics, Technion-Israel Institute of
 Technology, Technion City, 32000 Haifa, Israel}
\affiliation{Physics Department, Boston University, Boston, MA
 02215, USA}
\affiliation{Jefferson Laboratory of Physics, Harvard University,
 Cambridge, MA 02138, USA}

\author{Ryuichiro Kitano}
\affiliation{School of Natural Sciences, Institute for Advanced Study,
  Einstein Drive, Princeton, NJ 08540, USA}

\author{Hitoshi Murayama}
\affiliation{Department of Physics,
University of California, Berkeley, CA 94720, USA}
\affiliation{Theoretical Physics Group, Lawrence Berkeley National
Laboratory, Berkeley, CA 94720, USA\vspace*{7mm}}

\preprint{hep-ph/0504160}


\begin{abstract}
\vspace*{5mm}
Successful soft leptogenesis requires small $B$-terms for the
right-handed sneutrinos and a large CP violating phase between the $A$-
and $B$-terms. We show that this situation is realized naturally within
the framework of gauge mediated supersymmetry breaking.  The $A$-term is
dominated by contribution from gauge mediation, while supergravity
effects are more important for the $B$-term.  The different origins
naturally explain simultaneously the smallness of the $B$-term and the
large CP violating phase. The most stringent bounds on the model come from
the cosmological gravitino problem. We find a viable parameter region
with very light gravitino $m_{3/2} \lesssim 16\;\eV$, providing a
consistent framework for supersymmetry phenomenology, soft leptogenesis
and cosmology.
\end{abstract}

\vspace*{2cm}
\maketitle
 
\section{introduction}

Low-energy supersymmetry (SUSY) and leptogenesis~\cite{Fukugita:1986hr}
are both elegant ideas which account for theoretical and observational
insufficiencies of the standard model.
However, although each scenario is quite successful, their combination
is not necessarily good; there is basically no new positive effect in
SUSY leptogenesis that is absent in non-SUSY
leptogenesis~\cite{Plumacher:1997ru}.
Moreover, SUSY does introduce new problems.  In particular, the
gravitino, the superpartner of the graviton, causes serious cosmological
problems~\cite{Pagels:1981ke, Weinberg:1982zq, Khlopov:1984pf,
Kawasaki:2004yh}.
Big-bang nucleosynthesis puts a severe constraint on the reheating
temperature of the universe to avoid too much production of gravitinos
at high temperature.
A recent calculation gives an upper bound $T_{\rm RH} \lesssim
10^6\;\GeV$ for typical gravitino mass $m_{3/2} \sim
O(100\;\GeV)$~\cite{Kawasaki:2004yh}.  Such a bound is generally
inconsistent with the requirement of thermal leptogenesis $T_{RH} \gsim
M_N \gsim 10^{9}\;\GeV$~\cite{Giudice:2003jh}.\footnote{ Several ways
out had been pointed out in the literature, such as the right-handed
sneutrino condensate~\cite{Murayama:1993em}, heavy gravitino mass in
anomaly-mediated supersymmetry breaking~\cite{Ibe:2004tg}, and resonant
leptogenesis~\cite{Flanz:1994yx}.}

Recently, a new leptogenesis scenario, ``soft leptogenesis,'' was
proposed in which SUSY and SUSY breaking effects are positively
utilized~\cite{Grossman:2003jv, D'Ambrosio:2003wy}.  The presence of
the SUSY breaking terms enable the oscillation between right-handed
sneutrinos and their anti-particles to induce significant CP violation
in the sneutrino-decay processes.  Interestingly, successful soft
leptogenesis is achieved with relatively light right-handed
(s)neutrinos. Therefore, soft leptogenesis opens up a new possibility
of leptogenesis with low reheating temperature.

There are several conditions for soft leptogenesis to work. First, as
mentioned above, the right-handed neutrino has to be light,
$M_N \lesssim 10^8\;\GeV$, in order for the SUSY breaking effects to be
important. For fixed neutrino masses, this implies small Dirac Yukawa
coupling constants for neutrinos and thus the decay width of the
right-handed sneutrino is quite narrow.  In order for the oscillation
effect to be large, the mass splitting between the two mass eigenstates
of the right-handed sneutrinos cannot be too large or too small compared
to their decay width. Since the mass splitting is controlled by the SUSY
breaking $B$-parameter for the right-handed sneutrinos (see definitions
below), the narrow width implies that we need a small value for the
$B$-term. Finally, a large CP violating phase for the neutrino $B$-term
is necessary. This might be problematic because the phase of the Higgs
$B$-term is tightly constrained by low-energy experiments such as the
electric dipole moment of the electron and the
neutron~\cite{Ellis:1982tk}.

In particular, in gravity mediated SUSY breaking scenarios, small
$B$-term is difficult to realize since there is always a contribution of
the order of the gravitino mass under the condition that the
cosmological constant is canceled.  We need a certain level of
fine-tuning to realize $B \ll m_{3/2}$.  For small Yukawa coupling,
however, such smallness is stable under renormalization group equation
(RGE) evolution.  A model based on the assumption $B=0$ at tree level has
been considered in Ref.~\cite{Chun:2004eq}. Also, new ways of soft
leptogenesis without small $B$-terms which require very light
right-handed neutrinos, $M_N \lesssim 10^5\;\GeV$, have been discussed
in Ref.~\cite{Grossman:2004dz}.

In this paper we propose a natural framework that incorporates the above
conditions and satisfies all the phenomenological and cosmological
constraints such as CP violation in low-energy physics and the gravitino
problem.  The framework is quite simple; it is nothing but the gauge
mediated SUSY breaking scenario~\cite{Dine:1994vc,Giudice:1998bp}
without any additional structure.  In gauge mediation, the neutrino
$B$-term is generated by RGE running through the Yukawa interaction
because the right-handed neutrinos are gauge singlet. In that case, the
contribution from gravity mediation, which is of $O(m_{3/2})$, dominates
the $B$-term which generally carries an $O(1)$ CP violating phase. Since
in gauge-mediation models the gravitino mass is smaller than the SUSY
breaking scale, this is an ideal situation for soft leptogenesis.

This situation is perfectly consistent with bounds from low-energy
phenomenology. Gravity-mediation effects are completely negligible to
all other soft SUSY breaking terms, which are dominated by gauge
mediation for low enough SUSY breaking scales, i.e., small gravitino
mass parameters.  Therefore our assumption of the large CP violation
from gravity mediation does not pose new problems such as large flavor
changing neutral currents (FCNCs) or CP violation.  
In particular, for the most dangerous term for CP violation, the Higgs
$B$-term, the gauge-mediation contribution via RGE through the SU(2)$_L$
gauge interaction is much larger than the gravity-mediation effect.

We find a viable parameter region, $10^3\;\GeV \lesssim M_N \lesssim
10^6\;\GeV$ with $m_{3/2} \lesssim 16\;\eV$, where soft leptogenesis can
explain the baryon asymmetry of the universe without the gravitino
problem.  In gauge mediation, the gravitino is the lightest SUSY
particle and thus stable (we assume R-parity conservation). The
cosmological constraint comes from overproduction of the gravitinos at
high temperature~\cite{Moroi:1993mb, Pagels:1981ke} and also the
suppression of the matter power spectrum at small
scales~\cite{Viel:2005qj}.  However, these gravitino problems disappear
for $m_{3/2} \lesssim 16$~eV since, simply, the gravitino energy density
is small enough.

There is an interesting consistency of the scenario here.
Gauge-mediation models with such a light gravitino have been constructed
in Refs.~\cite{Gherghetta:2000qt, Goldberger:2002pc, Chacko:2005ra,
Izawa:1997hu}. In particular, the models in
Ref.~\cite{Gherghetta:2000qt, Goldberger:2002pc} are based on theories
with a warped extra dimension. Inspired by the AdS/CFT correspondence,
these theories are claimed to be equivalent to four dimensional models
with strong gauge dynamics.  In the extra-dimensional or the
conformal-field-theory setup, we are unable to discuss the conventional
leptogenesis because the cosmology would be quite different from the
four-dimensional or the weakly-coupled theory above temperatures of
order 10 to 100 TeV.  Since in our scenario the right-handed neutrinos
can be lighter than that, our model provides a consistent framework for
baryogenesis for such models.  Explicit models with strongly coupled
gauge theory are proposed in Ref.~\cite{Izawa:1997hu}. These models have
a weakly coupled description in high energy, and therefore both
conventional and soft leptogenesis are operative in different parameter
regions.

\section{soft leptogenesis}

We start by briefly reviewing the mechanism of soft
leptogenesis~\cite{Grossman:2003jv,D'Ambrosio:2003wy,Grossman:2004dz}.
For simplicity, we consider a one-generation toy model which consists
of the following superpotential
\beq
\label{suppot} 
W=\frac12 M_N NN + Y LNH.  
\eeq
Here $N$, $L$ and $H$ stand for, respectively, the gauge-singlet
right-handed neutrino, lepton doublet and up-type Higgs doublet chiral
superfields, and $M_N$ and $Y$ are the right-handed neutrino mass and
the Yukawa coupling constant, respectively.  Without SUSY breaking
terms, the mass and the width of the right-handed neutrino and sneutrino
are the same. Their mass is $M_N$ and their width is given by
\begin{eqnarray}
 \Gamma = \frac{Y^2 M_N}{4 \pi} = \frac{m M_N^2}{4 \pi v^2}\ ,
\qquad m \equiv {Y^2 v^2 \over M_N},
\end{eqnarray}
where $v \sim 174\;\GeV$ is the vacuum expectation value of the Higgs
field. 
We defined the neutrino-mass parameter $m$ which controls the efficiency
of the out-of-equilibrium decay and is naturally of the order of the
neutrino mass. Note that although it is identical to the neutrino mass
in this one-generation model, this is, in principle, an independent
parameter in a realistic three-generation model.  The soft SUSY breaking
terms relevant for soft leptogenesis are
\beq
\label{lsoft} 
{\cal L}_{\rm soft}=
\frac{B M_N}{2} \widetilde N \widetilde N
+A Y \widetilde L \widetilde N H + h.c.
\eeq 
This model has one physical CP violating phase
\beq
\label{cpvpha} 
\phi=\arg(A B^*).  
\eeq 

The soft SUSY breaking terms introduce mixing between the sneutrino
$\widetilde N$ and the anti-sneutrino $\widetilde{N}^\dagger$ in a
similar fashion to the $B^0 -\bar{B}^0$ and $K^0 - \bar{K}^0$
systems. The mass and width difference of the two sneutrino mass
eigenstates are given by
\beq \Delta m = |B|, \qquad \Delta \Gamma=
{2 |A| \Gamma \over M_N}.  
\eeq 
A non-vanishing CP violating phase $\phi$ induces CP violation in the
system.  The CP violation in the mixing generates the lepton-number
asymmetry in the final states of the $\widetilde N$ decay. This lepton
asymmetry is converted into the baryon asymmetry through the sphaleron
process.
The baryon to entropy ratio is given
by~\cite{D'Ambrosio:2003wy}:
\begin{eqnarray}
 \frac{n_B}{s} = - 10^{-3}\  \eta
\left[ \frac{4 \Gamma |B|}{4 |B|^2 + \Gamma^2} \right]
\frac{|A|}{M_N} \sin \phi\ .
\label{eq:asym}
\end{eqnarray}
The efficiency parameter $\eta$ slightly depends on the mechanism that
produces the right-handed sneutrinos. Assuming thermal production, the
value of $\eta$ is suppressed for small and large $m$ because of the
insufficient $\widetilde{N}$ production and strong washout effect,
respectively.
The maximum value is $O(0.1)$ for $m \sim
10^{-(3-4)}$~eV~\cite{D'Ambrosio:2003wy}.
{}From eq.~(\ref{eq:asym}) we
learn that the right-handed neutrino mass, $M_N$, cannot be too large
compared to the SUSY breaking scale because of the $|A|/M_N$ factor. Also,
the $B$-parameter should not be much larger or smaller than the
sneutrino width since otherwise the baryon asymmetry would be suppressed
by $\Gamma/|B|$ or $|B|/\Gamma$, respectively.
By fixing $m$, $A$, and
the phase $\phi$ such that eq.~(\ref{eq:asym}) takes its maximal value,
the above requirement gives a non-trivial constraint on the parameters
\cite{Grossman:2003jv,D'Ambrosio:2003wy}:
\beq 
 Y \lsim 10^{-4}, \qquad A \sim 10^2\;\GeV,\qquad
M_N \lsim 10^{8}\;\GeV,
\qquad B \lsim 1\;\GeV \qquad \phi \sim 1\ .
\eeq
Smaller values of $M_N$, corresponding to smaller $B$, are preferred to
avoid the gravitino constraint.
For example, if we are to avoid the gravitino problem by lowering $M_N$
to $10^6$~GeV, $B \lsim 1\;\MeV$ is necessary.
The value of $B$ is somewhat problematic as its naively expected value
is the weak scale, say, $B \sim 10^{2-3}\;\GeV$. It is our purpose to
find a framework that can generate such a small $B$ without affecting
the other parameters.

\section{Naturally small $B$-term}  
\label{sec:scenario}

Next we show that gauge-mediation provides a natural framework for soft
leptogenesis.  The basic idea we are proposing is as follows. Consider
gauge mediated SUSY breaking. This mechanism generates all the SUSY
breaking parameters via the standard model gauge
interactions~\cite{Dine:1994vc,Giudice:1998bp}.  Therefore, the $A$-term
in eq.~(\ref{lsoft}) is generated through the gauge interactions of $L$
and $H$. The $B$-term, on the other hand, remains very small because it
is gauge singlet and its Yukawa coupling is tiny.\footnote{Note that the
right-handed neutrinos do not have to be gauge neutral above the
messenger scale to avoid a large $B$-term.  In particular, SO(10)
unification is compatible with our framework.  } The dominant
contribution to the $B$-term, in this case, comes from the
gravity-mediation effect, which generates a $B$-term of the order of the
gravitino mass with a phase that is generally different from that of the
$A$-term.

We first estimate the size of the $A$- and $B$-terms in this scenario.
We consider a messenger scale, $M_{\rm msg}$, such that $M_N < M_{\rm
msg} \ll M_{\rm Pl}$.  At the messenger scale the gaugino masses, $M_i$,
and the scalar masses squared, $\widetilde{m}^2$, are generated at the
one- and two-loop levels, respectively, and thus $M_i$ and
$\widetilde{m}$ obtain $O(\alpha)$ contributions.
At this order the $A$- and $B$-terms are both vanishing. Non-vanishing
contributions at low energy are generated through renormalization-group
evolutions. The RGEs are given by
\begin{eqnarray}
 (4 \pi)^2 \frac{d}{d \log \mu} (AY) = 2 ( - g_Y^2 M_1 - 3 g_2^2 M_2 ) Y\ ,
\end{eqnarray}
\begin{eqnarray}
 (4 \pi)^2 \frac{d}{d \log \mu} (BM_N) = 8 M_N A Y^2\ ,
\end{eqnarray}
where $g_Y$ and $M_1$ ($g_2$ and $M_2$) are the gauge coupling constant
and the gaugino mass of the U(1)$_Y$ (SU(2)$_L$) group, respectively.
By integrating the above RGEs, we obtain the $A$- and $B$-parameters at
the scale $\mu = M_N$
\begin{eqnarray}
 A(M_N) = \frac{M_2}{g_2^2}
\left[
-\frac{5}{3} \beta_Y^{-1} ( g_Y^2(M_N) - g_Y^2(M_{\rm msg}))
-3 \beta_2^{-1} ( g_2^2(M_N) - g_2^2(M_{\rm msg}))
\right]\ ,
\label{eq:a-term}
\end{eqnarray}
\begin{eqnarray}
 B_{\rm gauge}(M_N) \simeq 
\frac{8}{(4 \pi)^2} A(M_N) Y^2 \log \frac{M_N}{M_{\rm msg}}\ ,
\label{eq:B-rge}
\end{eqnarray}
where the prefactor in eq.~(\ref{eq:a-term}), $M_2/g_2^2$, is an RGE
invariant quantity at one-loop level, and $\beta_Y$ and $\beta_2$ are
the beta-function coefficients. In the minimal SUSY standard model
(MSSM), those are given by $\beta_Y = 11$ and $\beta_2 = 1$.
Using the measured values at low energy, $g_i(M_Z)$, the gauge
coupling constants at $\mu = M_N$ and $M_{\rm msg}$ are given by
$g_i(M_Z)$:
\begin{eqnarray}
 \frac{1}{g_i^2 (\mu)} =  \frac{1}{g_i^2 (M_Z)} - 
 \frac{2 \beta_i}{(4 \pi)^2} 
\log \frac{\mu}{M_Z}\ ,
\label{eq:g-run}
\end{eqnarray}
where $i=Y,2$.

The crucial point to note from eqs.~(\ref{eq:a-term}), (\ref{eq:B-rge})
and (\ref{eq:g-run}) is that the $B$-term is very small.  An
$A$-term of the order of $g_2^2 M_2/(4 \pi)^2$  is obtained through
one-loop diagrams with gauge interactions. The $B$-term, however,
is further suppressed by a factor of $Y^2$ $(\lsim 10^{-8})$.
Although the smallness of the $B$-term is one of the requirements of
soft leptogenesis, this small $B$-term does not help us to get a viable
model. The main reason is that it is generated by loop diagrams through
the $A$-term and thus $\phi$, the physical CP violating phase, vanishes
at one-loop level.

Therefore, we need another mechanism that generates the $B$-term with
non-trivial CP violating phase.
This mechanism is already there in the form of gravity mediation.  As
long as $M_{\rm msg} \ll M_{\rm Pl}$, gravity-mediation contributions to
all other terms are negligibly small. Thus, it does not affect the good
features of gauge mediation, for example, the absence of FCNCs.
The gravity mediation is expected to generate
\beq
B_{\rm grav} \sim m_{3/2},
\label{eq:B-grav}
\eeq
with $m_{3/2}$ being the gravitino mass.
Since there is no connection between the phases of $A$ and $B_{\rm
grav}$, it is likely to be of order one, as desired.  The total $B$-term
is the sum of the two contributions, $B = B_{\rm grav} + B_{\rm
gauge}$. The phase $\phi$ is expected to be
\begin{eqnarray} \label{maxphi}
 \sin\phi \sim \left|
\frac{B_{\rm grav}}{B}
\right| \ .
\end{eqnarray}
In order to get a large CP violating phase we need the gravity-mediation
effect to dominate the $B$-term, $|B_{\rm grav}| \gsim |B_{\rm
gauge}|$. This is the case as long as $ m_{3/2} \gsim 10^{-3} m M_N /
v$.

\begin{figure}[t]
\centerline{\includegraphics[width=10cm]{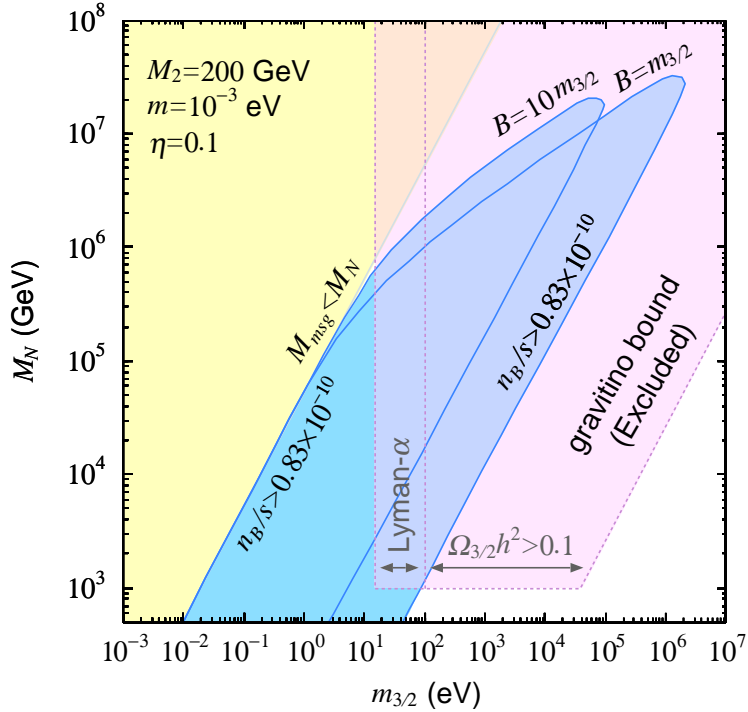}} \caption{The
region where we have successful baryogenesis is shown. We take
$M_2=200$~GeV, $m=10^{-3}\;\eV$, and $\eta = 0.1$. 
Contours of $n_B / s = 0.83 \times 10^{-10}$ are plotted for $B_{\rm
grav}=m_{3/2}$ and $10 m_{3/2}$ for the maximum CP phase. The baryon
asymmetry is larger inside the lines, where the correct value can be
obtained for smaller phases.
The messenger scale is lower than $M_N$ in the left-upper region. The
regions inside the dashed lines are excluded by the too large gravitino
energy density and the data from Lyman-$\alpha$ forest and WMAP.  }
\label{fig:m32-Mn}
\end{figure}

Now we can calculate the baryon asymmetry in this scenario.
Once we fix $m$, $m_{3/2}$, $M_2$ and $M_N$, we can estimate the baryon
asymmetry using eq.~(\ref{eq:asym}).
The messenger scale $M_{\rm msg}$ is obtained through the following
relations
\beq 
m_{3/2} = {F \over {\sqrt{3} M_{\rm Pl}}},
\qquad 
M_2= {g_2^2 N  \over (4 \pi)^2 } {F \over M_{\rm msg}} \ ,
\eeq 
where $F$ is the $F$-component of the hidden-sector field that breaks
supersymmetry, and $N$ is the Dynkin index of the messenger fields. In
our calculation we took $N=1$ which corresponds to a pair of ${\bf 5}$
and ${\bf \bar{5}}$ representation of the SU(5) group.
(Our results are not very sensitive to the choice of $N$ because the baryon
asymmetry depends on $N$ only through logarithmic functions.)
Then we can obtain the $A$-term by using eqs.~(\ref{eq:a-term}) and
(\ref{eq:g-run}), and the $B$-term as a sum of the contributions in
eqs.~(\ref{eq:B-rge}) and (\ref{eq:B-grav}).
The natural value of the phase $\phi$ is given in eq.~(\ref{maxphi}).
The Yukawa coupling constant in eq.~(\ref{eq:B-rge}) is estimated by
fixing $m$.

In our numerical calculation, we took $M_2 = 200\;\GeV$, $m =
10^{-3}\;\eV$, which corresponds to $\eta \sim 0.1$ (see
Ref.~\cite{D'Ambrosio:2003wy}), and two values for the $B$ parameter,
$B_{\rm grav}=m_{3/2}$ and $B_{\rm grav}=10 m_{3/2}$.  We found a
large region in the parameter space that gives successful leptogenesis
($m_{3/2} \lesssim 1$~MeV and $M_N \lesssim 10^7$~GeV).
We show in Fig.~\ref{fig:m32-Mn} the region in the
$m_{3/2}\;$--$\;M_N$ plane where large enough baryon asymmetry is
obtained.  The shaded regions inside the solid lines are allowed
corresponding to the values of $B$ as indicated.
The correct value of the baryon asymmetry is obtained on the lines when
the phase is maximum, i.e., $\sin \phi \sim B_{\rm grav}/B$, while 
smaller values are needed inside the lines.
In the left-upper region, the messenger scale is lower than the
right-handed neutrino mass scale where gauge mediation does not generate
the needed $A$-term.
The region with $M_N \ll 1\;\TeV$ is not allowed since in that case the
sphaleron process is not active when the right-handed sneutrinos decay.

We checked the sensitivity of the allowed region to the input
parameters.  The allowed region is not very sensitive to $M_2$. On the
other hand, it is rather sensitive to the variation of $m$ mainly due
to the fact that then $\eta$ become smaller.

\section{constraints from gravitino cosmology}

We consider the constraints on the model parameters from gravitino
cosmology.
In gauge-mediation models, the gravitino is the lightest SUSY particle
and thus it is stable. Therefore the gravitinos generated at high
temperature contribute to the matter density of the universe.
In order for the gravitino energy density $\Omega_{3/2} h^2$ not to
exceed the measured (non-baryonic) matter density of the universe,
$\Omega_{\rm DM} h^2 = 0.11$, the reheating temperature is constrained
to be~\cite{Moroi:1993mb, Pagels:1981ke}:
\beq \label{eq:closebound}
T_{RH} \lesssim \cases { 
\displaystyle
1\;\TeV\; 
\left[ {\displaystyle m_{3/2} \over \displaystyle 100\;\keV} \right]
\left[ {\displaystyle M_3 \over { \displaystyle 1\;\TeV}} \right]^{-2}
& for $m_{3/2} \gsim 100\;\keV$\,, \cr 
1\;\TeV\;
& for $100\;\keV \gsim m_{3/2} \gsim 100\;\eV$\,, \cr 
\mbox{no bound} & for $m_{3/2} \lesssim 100\;\eV$\,,}\  
\eeq 
where $M_3$ is the gluino mass.
Note that there is no constraint for $m_{3/2} \lesssim
100\;\eV$. 
This is because, for such light gravitinos, the goldstino component of
the gravitino has large interaction rate with the MSSM particles, and
therefore the gravitino can thermalize at high temperature. The number
density in this case is just the equilibrium value and thus the energy
density is approximately given by $\Omega_{3/2} h^2 \sim 0.1
(m_{3/2}/100\;\eV)$ with taking into account the dilution effect by the
decoupling of heavy particles.

In this light gravitino region, we have another constraint from the
matter power spectrum obtained by the Lyman-$\alpha$ forest and the WMAP
data~\cite{Viel:2005qj}.  Since the gravitinos are warm (the
free-streaming scale is comparable to galaxy scales) once they are
thermalized, they would smear out the density perturbation at small
scales if the gravitino contributes significantly to the matter density
even though it is subdominant.
This requires small $m_{3/2}$ such that the gravitino energy density is
less significant.
The excluded region is~\cite{Viel:2005qj}
\beq \label{eq:forestbound}
16~\eV \lesssim m_{3/2} \lesssim 100\;\eV\ .
\eeq

Assuming that right-handed sneutrinos are thermally produced, the
reheating temperature has to be higher than $M_N$. Therefore
eq.~(\ref{eq:closebound}) gives a non-trivial constraint on the
parameters. On the other hand, the excluded region in
eq.~(\ref{eq:forestbound}) applies as long as the gravitinos are once
thermalized.
The bounds from eqs.~(\ref{eq:closebound}) and (\ref{eq:forestbound})
are superimposed on the allowed region in Fig.~{\ref{fig:m32-Mn}}. 
We see that we still have an allowed region for very light gravitinos
$m_{3/2} \lesssim 16~\eV$ and $10^3~\GeV \lesssim M_N \lesssim
10^6~\GeV$.

Once we specify a mechanism for non-thermal production of the
right-handed sneutrinos, the condition of $ T_{RH} > M_N $ is relaxed
and therefore the region where $m_{3/2} \gtrsim 100~\keV$ revives.
The region $16~\eV \lesssim m_{3/2} \lesssim 100~\keV$ is still excluded
since the bound is applicable as long as the gravitinos are thermalized.
Our scenario needs the sphaleron process to be active. That requires
temperatures higher than about $300\;\GeV$ where some of the SUSY particles
are in thermal equilibrium with reasonable SUSY-breaking scales.
The gravitinos are thermalized in that circumstance, and therefore the
region is excluded regardless of the mechanism of the right-handed
sneutrino production.
Examples of the non-thermal production are direct production from
inflaton decay and non-thermal production from coherent oscillation of
$\widetilde{N}$.
Such possibilities are worth further studies.

\section{Discussions and conclusions}
\label{sec:conclusions}

We considered soft leptogenesis in gauge mediated SUSY breaking
scenario. We found that the needed small $B$-term and large CP
violating phase are naturally obtained through gravity-mediation
effects.  We found a region in the parameter space with light
right-handed neutrino $10^3~\GeV \lesssim M_N \lesssim 10^6\;\GeV$ and
light gravitino $m_{3/2} \lesssim 16\;\eV$, where the model observe
all experimental and cosmological constraints.

The large CP phase in the neutrino $B$-term would not imply large
phases in other soft terms, which would conflict with the low energy
data.
The source of the phase, the supergravity effect, is important for the
terms which only involve gauge-singlet fields.
For all other terms the gauge-mediation contributions overwhelm the tiny
gravity-mediation effect, realizing an ideal situation for soft
leptogenesis and low energy phenomenology.

Gauge mediated SUSY breaking with $m_{3/2} \lesssim 16\;\eV$ is a very
interesting possibility among various SUSY breaking scenarios because it
completely avoids the FCNC/CP and cosmological gravitino problems.
If we confirm the spectrum pattern of gauge mediation at the LHC/ILC
experiments, this soft-leptogenesis scenario becomes an outstanding
possibility for baryogenesis among various mechanisms.
The low SUSY breaking scale, corresponding to the
light gravitino, requires some models with strong dynamics or
(probably equivalently) extra dimensions. In the extra-dimensional
models, where the high-temperature physics is unclear, our scenario
provides a simple and natural mechanism of baryogenesis at low
temperature.

\acknowledgments 
We thank Esteban Roulet and Yossi Nir for comments.
We also thank the Aspen Center for Physics where this work was
initiated. RK thanks the hospitality of the High Energy Theory Group at
Harvard University.
This work was supported by funds from the
Institute for Advanced Study.
The work of HM was supported in part by the DOE under contract
DE-AC03-76SF00098 and in part by NSF grant PHY-0098840.


\end{document}